# Predictability and epidemic pathways in global outbreaks of infectious diseases: the SARS case study

Vittoria Colizza[1,*], Alain Barrat[1,2], Marc Barthélemy[3] and Alessandro Vespignani[4,5]

[1]Complex Networks Lagrange Laboratory (CNLL), Institute for Scientific Interchange (ISI) Foundation, Turin, Italy

[2]Unité Mixte de Recherche du CNRS UMR 8627, Bâtiment 210, Univ Paris-Sud,  F-91405 Orsay, France

[3]CEA-DIF Centre d'Etudes de Bruyères-Le-Châtel, BP12, F- 91680, France

[4]School of Informatics and Center for Biocomplexity, Indiana University,  Bloomington, IN 47401, USA

[5]Institute for Scientific Interchange (ISI) Foundation, Turin, Italy

*Corresponding author: vcolizza@isi.it

# Abstract

### Background

The global spread of the severe acute respiratory syndrome (SARS) epidemic has clearly shown the importance of considering the long-range transportation networks in the understanding of emerging diseases outbreaks. The introduction of extensive transportation data sets is therefore an important step in order to develop epidemic models endowed with realism.

### Methods

We develop a general stochastic meta-population model that incorporates actual travel and census data among 3 100 urban areas in 220 countries. The model allows probabilistic predictions on the likelihood of country outbreaks and their magnitude. The level of predictability offered by the model can be quantitatively analyzed and related to the appearance of robust epidemic pathways that represent the most probable routes for the spread of the disease.

### Results

In order to assess the predictive power of the model, the case study of the global spread of SARS is considered. The disease parameter values and initial conditions used in the model are evaluated from empirical data for Hong Kong. The outbreak likelihood for specific countries is evaluated along with the emerging epidemic pathways. Simulation results are in agreement with the empirical data of the SARS worldwide epidemic.

**Conclusions**

The presented computational approach shows that the integration of long-range mobility and demographic data provides epidemic models with a predictive power that can be consistently tested and theoretically motivated. This computational strategy can be therefore considered as a general tool in the analysis and forecast of the global spreading of emerging diseases and in the definition of containment policies aimed at reducing the effects of potentially catastrophic outbreaks.

# Background

The outbreak of severe acute respiratory syndrome (SARS) in 2002–2003 represented a serious public health threat to the international community. Its rapid spread to regions far away from the initial outbreak created great concern for the potential ability of the virus to affect a large number of countries and required a coordinated effort aimed at its containment [1]. Most importantly, it clearly pointed out that people's mobility and traveling along commercial airline routes is the major channel for emerging disease propagation at the global scale. Spatio-temporal structures of human movements thus need to be considered for a global analysis of epidemic outbreaks [2], as for example in [3], which incorporates the airline network structure of the largest 500 airports of the world.

In this article, we present a stochastic meta-population epidemic model, based on the extension of the deterministic modeling approach to global epidemic diffusion [4,5], for the study of the worldwide spread of emerging diseases that includes the complete International Air Transport Association (IATA) commercial airline traffic database associated with urban areas census information [6,7]. Once the disease parameters are determined, no free adjustable parameters are left in the model. A toolkit of specific indicators that consider the stochastic nature of the process is introduced to provide risk analysis scenarios and to assess the reliability of epidemic forecasts. In particular, the predictive power of the model is linked to the emergence of epidemic propagation pathways related to the complex properties of the transportation network. The SARS epidemic is used as a case study to assess the model effectiveness and accuracy against real data. The model considers disease parameters estimated from the Hong Kong outbreak in a way consistent with the global nature of the meta-population model by including the impact of infectious individuals traveling in and out of the city. The temporal and geographic pattern of the disease is analyzed, and the proposed toolkit of epidemic indicators is tested against empirical data.

# Methods

We adopt a global stochastic meta-population model that considers a set of coupled epidemic transmission models. The approach is in the same spirit as the deterministic models used for the global spread of infectious diseases and their successive stochastic generalizations [3,6,7], where each compartmental model represents the evolution of the epidemic within one urban area, and the models are coupled by air travel. The air travel data from the IATA [8] database is included in the model and determines the traveling probabilities. It includes the 3 100 largest commercial airports around the globe and 17 182 connections among them, accounting for more than 99% of the total worldwide traffic. Each airport is surrounded by the corresponding urban area whose population is assumed to be homogeneously mixed for the disease dynamics. Population data is collected from several census databases (see [6,7] for more specific details). The model is fully stochastic and takes into account the discrete nature of individuals both in the travel coupling and in the compartment transitions. The transmission model within each urban area follows a compartmentalization specific to the disease under study. For instance, in the case of a simple Susceptible-Infected-Recovered (SIR) model, the population $N_j$ of a city $j$ is subdivided into susceptible, infectious and recovered individuals, so that $N_j = S_j(t)+I_j(t)+R_j(t)$ where $S_j(t)$, $I_j(t)$ and $R_j(t)$ represent the number of individuals in the corresponding compartments at time $t$. In order to consider the discrete nature of the individuals in the stochastic evolution of the infection dynamics, we describe the disease propagation inside each urban area by introducing binomial and multinomial processes. Two kinds of processes are considered in the infection dynamics: the contagion process (e.g. the generation of new infectious through the transmission of the disease from infectious individuals to susceptibles) and the transition of individuals from one compartment to another (i.e. from infectious to recovered). In the first class of processes it is assumed that each susceptible in city $j$ will be infected by the contact with an infectious individual with rate $\beta I_j(t)/N_j$, where $\beta$ is the transmission rate of the disease. The number of new infections generated in city $j$ is extracted from a binomial distribution with probability $\beta I_j(t)\Delta t/N_j$ and number of trials $S_j(t,)$, where $\Delta t$ is the considered time scale interval. The second class describes a transition process, where the number of individuals changing compartment: e.g. in the SIR model for the city $j$: $I_j \rightarrow R_j$ with rate $\mu$ - is extracted from a binomial distribution with a probability given by the rate of transition (in the previous example $\mu$) and number of trials given by the number of individuals in the compartment at time t (in the previous example $I_j(t)$).

Changing from a basic SIR model to a refined compartmentalization, additional processes ought to be taken into account, as the possibility of having more than one compartment able to transmit the infection, due e.g. to the non-perfect isolation of quarantined individuals. In the case of SARS, which will be addressed in the following section, the infection dynamics includes the specific characteristics of the disease under study, such as latency, hospitalization, patient isolation, and fatality rate [9,10,11]. Figure 1 illustrates a schematic representation of the compartmentalization adopted for the SARS case study, whereas the details of the stochastic discrete evolution followed by this specific compartmentalization are described in Additional file 1.

Each compartmental model in a given urban area is then coupled to the compartmental models of other urban areas via a travel stochastic operator that identifies the number of individuals in each compartment traveling from the urban area $i$ to the urban area $j$. The number of passengers in the compartment $X$ traveling from a city $i$ to a city $j$ is an integer random variable, in that each of the $X_i$ potential travellers has a probability $p_{ij} = w_{ij}/N_i$ to go from $i$ to $j$ where $w_{ij}$ is the traffic, according to the data, on a given connection in the considered time scale and $N_i$ is the urban area population. In each city $I$, the numbers of passengers traveling on each connection at time $t$ define a set of stochastic variables that follows a multinomial distribution. In addition, other routing constraints and two legs travels can be considered. A detailed mathematical description of the traveling coupling is reported in [6, 7,12].

The defined model considers stochastic fluctuations both in the individual compartmental transitions and in the traveling events. This implies that in principle each model realization, even with the same initial conditions, may be different from all the others. In this context, the comparison of a single realization of the model with the real evolution of the disease may be very misleading. Similarly, the mere comparison of the number of cases obtained in each country averaged over several realizations with the actual number of cases occurred is a poor indicator of the reliability of the achieved prediction. Indeed in many cases the average would include a large number of occurrences with no outbreaks in a variety of countries. It is therefore crucial to distinguish in each country (or to a higher degree of resolution, in each urban area) the non-outbreak from the outbreak realizations and evaluate the number of cases conditionally to the occurrence of the latter events. For this reason, we define in the following a set of indicators and analysis tools that can be used to provide scenarios forecast and real world data comparison.

## Outbreak likelihood and magnitude

The likelihood to experience an outbreak can be provided by analyzing different stochastic occurrences of the epidemic with the same initial conditions, and by evaluating the probability that the infection will reach a given country. In the following we will consider statistics over $10^3$ different realizations of the stochastic noise, and define the probability of outbreak in each country as the fraction of realizations that produced a positive number of cases within the country. This allows for the identification of areas at risk of infection, with a corresponding quantitative measure expressed by the outbreak probability. A more quantitative analysis is obtained by inspecting the predicted cumulative number of cases for each country, conditional to the occurrence of an outbreak in the country. The outbreak likelihood and magnitude analysis can be broken down at the level of single urban areas. In the following section we present an example of the results available at this resolution scale.

## Predictability and epidemic pathways

The very high potential value of forecasting tools, in a planning perspective against emerging infectious diseases, points to the necessity of assessing the accuracy of such epidemic forecasts with respect to the various stochastic elements present in the process. Indeed, the present computational approach provides meaningful predictions only if all stochastic realizations of the epidemic, with the same initial conditions and parameters, are somehow similar in intensity, locations and time evolution. The airline network structure explicitly incorporated into the model is composed by more than 17 000 different connections among 3 100 cities. Such a large number of connections produce a huge amount of possible different paths available for the infection to spread throughout the world. This in principle could easily result in a set of simulated epidemic outbreaks that are very different one from the other – though starting from the same initial conditions – thus leading to a poor predictive power for the computational model. By contrast, while the airline network topology tends to lower the predictability of the disease evolution, the heterogeneity of the passenger volume on the various connections defines specific diffusion channels on the high traffic routes. Ultimately, the degree of predictability is determined by the competing effects of connectivity and traffic heterogeneities [6,7] once the initial starting conditions and parameters of the disease are fixed [13]. In order to measure quantitatively the stochastic variability in the computational model of a specific disease we follow [6,7] and consider an overlap function $\Theta(t)$, which describes the similarity of two different outbreak realizations starting from the same initial conditions. The overlap function measures the similarity of two different realizations of the epidemic outbreak by comparing the evolution in time of the number of *active* individuals $A_j(t)$ in each urban area $j$, defined as those individuals carrying the infection. Each outbreak starting from the same initial conditions is characterized by a vector $\vec{\pi}(t)$ whose $j$-th component represents the probability that an active individual is in city $j$ at time $t$, $\pi_j(t) = A_j(t) / \sum_l A_l$. Following [6,7], we compute the statistical similarity between two different realizations characterized by the vectors $\vec{\pi}^I(t)$ and $\vec{\pi}^{II}(t)$, respectively, by considering the Hellinger affinity defined as $sim(\vec{\pi}^I, \vec{\pi}^{II}) = \sum_j \sqrt{\pi_j^I \pi_j^{II}}$. This normalized measure is invariant under a rescaling of the vectors $\vec{\pi}^I(t)$ and $\vec{\pi}^{II}(t)$ by a constant factor. Therefore, we introduce also the similarity between the worldwide epidemic prevalence obtained in the two realizations: $sim(\vec{a}^I, \vec{a}^{II})$, where $\vec{a}^{I(II)} = (a^{I(II)}, 1 - a^{I(II)})$ and $a(t) = \sum_j A_j / \aleph$, with $\aleph = \sum_j N_j$ being the world population. The overlap function $\Theta(t)$ is thus defined as:

$$\Theta(t) = sim\left(\vec{a}^I(t), \vec{a}^{II}(t)\right) sim\left(\vec{\pi}^I(t), \vec{\pi}^{II}(t)\right).$$

The overlap $\Theta(t)$ assumes values between 0 and 1, being equal to 0 if at time $t$ the two epidemic patterns do not share any common infected city, and equal to 1 if at time $t$ the two realizations are identical. The more an outbreak is predictable, the more likely the two realizations will be similar, leading to a high value of the overlap function. In view of the strong fluctuations inherent to the infection process and the movement of individuals, the presence of an appreciable overlap can be possible only in the presence of a robust mechanism driving the disease propagation and leading to the emergence of epidemic pathways, i.e. preferential channels along which the epidemic will more likely spread [6,7]. These pathways on their turn may find their origin in the large heterogeneities encountered in the traffic volume – ranging from a few passengers to $10^6$ passengers per year – associated with the air travel connections. In order to pinpoint the presence of epidemic pathways, starting from identical initial conditions, one can simulate different outbreaks subject to different realizations of the stochastic noise and obtain the time evolution of the epidemic in each urban area as described in the main text. During the simulations, one observes the propagation of the virus from one country to the other by means of the air travel and thus monitors the path followed by the infection at the country level. At each outbreak realization, it is possible to identify for each country $C_i$ the country $C_j$ origin of the infection and construct the graph of virus propagation; namely, if a latent or an infectious individuals travels from $C_j$ to $C_i$ and causes an outbreak in the country $C_i$ – not yet infected – a directed link from $C_j$ to $C_i$ is created with weight equal to 1. Once the origin of infection for $C_i$ has been identified, the following multiple introductions in $C_i$ are not considered as we are only interested in the path followed by the disease in infecting a geographical region not yet infected. After a statistically significant number of realizations, a directed weighted network is obtained in which the direction of a link indicates the direction of the virus diffusion and the weight represents the number of times this flow has been observed out of $n$ realizations. For each country $C_i$ we renormalize to 1 the sum of the weights on all incoming links, in order to define the probability of infection on each flow. The network of epidemic pathways is then pruned by deleting all directed links having an occurrence probability less than a given threshold, in order to clearly identify the major pathways along which the epidemic will spread. This information identifies for each country the possible origins of infection and provides a quantitative estimation of the probability of receiving the infection from each identified origin. It is therefore information of crucial importance for the development and assessment of preparation plans of single countries. Travel advisories or limitations and medical screenings at the ports of entry – such as those put in place during SARS epidemic – might well strongly benefit from the analysis and identification of such epidemic pathways.

## Results

As a concrete example of the previous modeling approach we analyze the specific case study of the SARS epidemic. Several mathematical models have been developed since the SARS coronavirus was identified (see [14] and references therein). Many of these approaches focused on localized communities, such as generic hospital populations [15], specific cities or small regions [9,10,16-27], whereas few others considered the role of global travel [3,28]. In particular, the estimates of key epidemiological parameters are traditionally obtained from fitting local models (i.e. on the scale of cities, regions) to empirical data. Such approaches assume closed boundary conditions for the region under consideration, neglecting possible movements of individuals in and out of the region. In the following we use the global computational model defined in the previous section to simulate synthetic SARS outbreak on the worldwide scale and compare with the empirical data from the real world occurrence. The compartmentalization and the disease parameter values considered in the model are chosen according to previous studies [1,11,29]. It is worth stressing however that the compartmentalization used (see Figure 1), while borrowed from the most authoritative references on the SARS outbreak, is suffering from the approximations due to the lack of information on the social specific structure and heterogeneity that might be crucial for a full understanding of the disease. Parameters are kept constant throughout the evolution of the disease, except for the hospitalization rate $\mu^{-1}$ and for the scaling of the transmission rate $\beta$, which follow a three steps function as in [11] and [9], respectively (Table 1). These steps correspond to the implementation in Hong Kong of containment measures and advisories effectively reducing the transmission rate [9,11]. It is reasonable to consider a certain reaction time delay taking place in each country from the detection of the first case to the implementation of the policies aimed at reducing the transmission rate on a national scale. In the following, we report the results for an intervention delay of 1 week. The Additional file reports the results for immediate reaction (no delay) and 2 weeks delay.

Initial conditions are based on available evidence on the early stages of the outbreak and assume as index patient the first case detected out of mainland China, who arrived in Hong Kong on 21 February 2003 [30]. Simulations are seeded in Hong Kong $T_0$ days after 21 February with the index patient and $L(t = 0)$ initial latent. This allows the effective consideration of the observed super-spreading events [31-34] and multiple transmission before the index patient was hospitalized [9]. The complete time frame under study is from $T_0$ days after 21 February to 11 July 2003, date corresponding to the last daily update by the World Health Organization (WHO) on the cumulative number of reported probable cases of SARS [35].

The values of the transmission rate $\beta$, of the initial number of latent individuals $L(t = 0)$, and of the initial delay of $T_0$ days are determined through a least square fit procedure to optimize the agreement of the stochastic simulation results with Hong Kong data. The advantage with respect to previous approaches is that no closed boundaries are imposed on Hong Kong, allowing for the mobility of individuals traveling in the city and for a decrease of the pool of infectious individuals who leave the city by means of air travel. The optimization gives the following baseline values: $\beta = 0.57\,[0.56\text{–}0.59]$, $L(t = 0) = 10\,[8\text{–}11]$, $T_0 = 3$ days, where the errors reported for $\beta$ and $L(t = 0)$ correspond to a relative variation of 10% in the least square value from the minimum, once the offset is set to its optimal value. The obtained value of the reproductive number – $R_0 = 2.76$ – is in agreement with previous estimates [9]. We also tested different initial conditions that do not effectively incorporate super-spreading events with no substantial changes in the results.

## Outbreak likelihood

In Figure 2 we represent on a map the countries that are more likely to be infected with a color code, ranging from gray, signaling low outbreak probability, to red for a high probability of experiencing an outbreak. It represents a quantitative indication of the risk to which each country would be exposed in presence of a SARS-like infectious disease in which the same containment measures are implemented. It therefore provides a starting point for the development of appropriate intervention scenarios aimed at reducing that risk. The map readily identifies geographical areas with an appreciable likelihood for an outbreak. In particular, many countries of South-East Asia display a large probability of outbreak, as could be expected from their vicinity from the initial seed. The fact that Western Europe and North America also suffer outbreaks in most realizations illustrates the role of large air traffic in the propagation events. Figure 3 maps the outbreak likelihood at a higher resolution scale, i.e. at the level of urban areas, showing the expected situation in Canada. It is worth noting that despite the large number of airports in the country, only very few areas display a significant probability of outbreak, the two largest values corresponding to the actual outbreaks experienced in Toronto and Vancouver.

To proceed further in the comparison with empirical data, we group countries in two categories according to a risk threshold in the outbreak occurrence probability. The no-risk countries are those where the probability of outbreak is lower than the risk threshold. In any other situation the country is defined at risk. In the following we set a risk threshold of 20%. Small variations of the risk threshold do not alter substantially the obtained results. In particular we show in Additional file 1 the results for the case of a 10%, and 30% risk threshold. The results do not differ considerably from those reported for the 20% risk threshold. Obviously, progressively larger values of the risk threshold leads to less significant results, and a risk threshold of 40–50% is not providing valuable information as defining a not at risk a country with a 45% outbreak probability would be quite unreasonable. In Figure 4 we represent the comparison between data reported by WHO on 11 July and the results obtained from the numerical simulations for the same date and the 20% risk threshold. The report for 11 July, cross-checked with the WHO final summary obtained after laboratory tests, was of 28 infected countries, spread on different continents [35]. Figure 4 shows in red the agreement of our results with the empirical data, with full color for countries that our simulations declare at risk and are found to be infected according to the WHO report (correct prediction of outbreak), and striped pattern for countries that are not infected according to official reports and which we classify as no-risk countries. Countries that belong to a class that is not in agreement with the data are represented in green. The map shows a very good agreement of forecasts with official updates: the simulations capture the worldwide spread of SARS by recovering with high success rate countries that either were infected or remained uninfected. Out of a total number of 220 countries in the world, simulations are able to predict correctly 23 of the 28 infected countries by 11 July. Five countries are not classified at risk in the simulations while they reported probable cases – Mongolia, Sweden, Kuwait, Ireland, Romania – and 10 countries are considered at risk while no official update ever reported any case (Table 2). In other words, the model provides a correct classification for 205 out of 220 countries. Additional analysis shows that the agreement of our forecasts with empirical data extends over the whole duration of SARS outbreak, with a maximal error of about 7% only (wrong classification for 15 countries out of 220).

**Outbreak magnitude**

A more quantitative analysis is obtained by comparing the predicted cumulative number of cases for each country, conditional to the occurrence of an outbreak in the country, with the corresponding empirical data [35]. The forecasts on the cumulative number of cases are reported in Figure 5 with a box plot representing median, quartiles and 90% CI and compared to the empirical data (red symbols). All countries listed in the last WHO report are shown for analysis, regardless of their predicted probability of outbreak. Even for the five countries not classified at risk it is indeed possible to test simulations against empirical data by considering the cumulative number of cases conditional to the presence of an outbreak. Panel (A) shows the countries for which the empirical data lie inside the fluctuations associated with the model forecasts. For 22 countries, out of the 28 of the final report, the model is able to provide results where the actual number of cases falls within the error bars of the statistical predictions. The best agreement is obtained for Hong Kong, as expected, as the initial conditions were fitted to Hong Kong data. It is worth to note that all five countries wrongly predicted as no-risk show statistics conditional to the outbreak in agreement with empirical data (see inset of panel (A)). Panels (B) and (C) correspond to the predictions for which the observed data lie outside the range of fluctuations of the model. Five countries fall in this category, two of which – Thailand and Republic of Korea – report fluctuations very close to the empirical data. While at first sight the error bars associated to the model predictions might appear large and the results of panels (B) and (C) of Figure 5 for large outbreaks quite far from the real occurrence, it is worth stressing that the predicted numbers of cases are strikingly in the correct ballpark if we consider that we are dealing with a few hundred cases over the total country population, typically of the order of several millions. Finally, Table 2 reports data for the 10 countries predicted at risk by simulations but where no infection from SARS was reported. Japan represents an outlier: a large number of cases is predicted by the numerical simulations (the median being 83). Remarkably however, the predicted number of cases remains small (median at most equal to six cases) in the other nine countries. Forecasts reported in Figures 2–5 and Table 2, though with some deviations from empirical data, provide quantitative evidence in agreement with previous findings [3] that air travel represents a crucial ingredient for describing SARS propagation on a global scale. In the Additional file 1 we report the outbreak magnitude in the case of no delay and 2 weeks delay in the reduction of the transmission rate since the detection of the first case in each country. It is possible to observe that for increasing delays the simulations results increase their accuracy for countries experiencing large outbreaks such as Canada and Singapore. The results thus show that the delay has an impact in determining the extent of the outbreak in certain countries. It is therefore important to stress that it is very likely different delays have to be considered in each country depending on the local health infrastructure.

**Overlap and epidemic pathways**

In order to test the predictability inherent to the model in the case of the SARS case study, Figure 6 shows the overlap as defined in the Methods section as a function of time. The average value is displayed together with the 95% confidence interval. The overlap starts from a value equal to 1, as all stochastic realizations share the same initial conditions, and decreases monotonically with time. However, $\Theta(t)$ assumes values larger than 0.8 in the time window investigated, confirming the relatively strong computational reproducibility of the synthetic SARS outbreak. The simulated disease seems indeed to follow a very similar evolution at each realization of the process. As discussed in the previous section, the origin of such reproducibility lies in the emergence of epidemic pathways, i.e. preferential channels along which the epidemic will more likely spread [6,7]. In order to identify these pathways, we monitor the spreading path followed by the virus in $10^3$ outbreaks starting from the same initial conditions. More precisely, starting from Hong Kong, we follow the propagation of the virus and identify for each infected country $C_i$ the country $C_j$ where the infection came from, thus defining a probability of origin of infection for each country. Results are reported in Figure 7 where the epidemic pathways are represented by arrows whose thickness accounts for the probability of infection. Almost every country in which at least one case was detected in our simulations received the infection most likely directly from Hong Kong, with probabilities ranging from 32% for Italy, to 99% for Taiwan. Spain is the only exception as it belongs to a second level of infection from the seed and is predicted to receive the infection from other European countries (see the bottom left panel of Figure 7). While this is not a completely surprising results in the case of an airport hub such Hong Kong, it is worth noting that results in Figure 7 do not represent the distribution of infecting paths going out of a given country but rather the probability associated to infecting paths entering in a given country. In the case of a different starting infected city, epidemic pathways can be highly nontrivial and therefore informative in the deployment of monitoring and control resources.

# Discussion

While the results shown in Figures 2–5 indicate that computational models can attain a predictive power, the present case study for the SARS epidemic is still subject to approximations and assumptions and very accurate predictions need the introduction of much more detail into the computational scheme. Taiwan is the prominent example in which the quantitative magnitude of the outbreak is not well predicted. Indeed, the model still lacks many features that are likely responsible for the deviations of simulation results from the official reports. Population heterogeneity in terms of travel frequency, obviously related to wealth distribution, is not considered. Specific features relative to single countries (such as health care systems, specific control strategies, travel screening etc.) are not taken into account by our stochastic model. We also do not consider variations in the virus transmissibility among infected individuals in each compartment. Similarly, we account only in an effective way for the occurrence of superspreading events and outbreak diversity in the initial stage of SARS transmission [33,34,36]. By contrast, the model readily allows the integration of the heterogeneity shown by SARS in the way it affected different countries and which represented a very peculiar feature of the virus. Finally, the inclusion of other transportation systems is likely to have an impact in contiguous geographical regions.

It is worth also noting, however, that the countries for which forecasts underestimate the empirical data showed some peculiarities in the evolution of the SARS spread. Taiwan for instance experienced an anomalous outbreak explosion after a temporary failure in the infection containment procedures in a single hospital [37]. Moreover, it ought to be considered that the situation in mainland China is not trivially reproducible, due to the lack of available information on the actual initial conditions of the spread. Results for China are therefore not reported in the charts of Figure 5, as numerical simulations seeded in Hong Kong are likely not able to describe the outbreak occurred in that country. This fact is expected to have an impact especially in South-East Asian countries – such as Taiwan, Singapore and Vietnam – that have airline connections with large traffic towards China.

## Conclusions

The computational approach presented here is the largest scale epidemic model at the worldwide level. Its good agreement with historical data of the SARS epidemic suggests that the transportation and census data used here are the basic ingredients for the forecast and analysis of emerging disease spreading at the global level. A more detailed version of the model including the interplay of different transportation systems, information about the specific conditions experienced by each country and a refined compartmentalization to include variations in the susceptibility and heterogeneity in the infectiousness would clearly represent a further improvement in the *a posteriori* analysis of epidemic outbreaks. In the case of a new emergent global epidemic the computational approach could be useful in drawing possible scenarios for the epidemic evolution. Though the initial conditions and the disease parameters will be unknown before the disease has already spread to a few countries, the computational approach would however allow in a short time the exploration of a wide range of values for the basic parameters and initial conditions, providing extensive data on the worst and best case scenarios as well as likelihood intervals, to the benefit of decision makers. In general, the encouraging results achieved with the present level of details introduced in the modeling schemes suggests that large scale computational approaches can be a useful predictive tool to assess risk management and preparedness plans for future emerging diseases and in understanding space-time variations of outbreak occurrences.

## Competing interests

The authors declare that they have no competing interests.

## Authors' contributions

All authors conceived the study, collected data and performed experiments for the study, analyzed results and contributed to writing the paper. All authors read and approved the final manuscript.

# Acknowledgements

The authors thank the International Air Transport Association for making the commercial airline database available. AB and AV are partially funded by the European Commission- contract 001907 (DELIS). AV is partially funded by the NSF award IIS-0513650.

# Figure Legends

## Figure 1 - Flow diagram of the transmission model

The population of each city is classified into seven different compartments, namely susceptible ($S$), latent ($L$), infectious ($I$), hospitalized who either recover ($H_R$) or die ($H_D$), dead ($D$) and recovered ($R$) individuals. We assume that hospitalized as well as infectious individuals are able to transmit the infection, given the large percentage of the cases among health care workers [37-39]. The actual efficiency of hospital isolation procedures is modeled through a reduction of the transmission rate $\beta$ by a factor $r_\beta$ for hospitalized patients, with $r_\beta$ = 20% as estimated for the early stage of the epidemic in Hong Kong [9]. The infectiousness of patients in the compartments $H_R$ and $H_D$ are assumed to be equal (although this assumption can easily be changed in the model). Susceptible individuals exposed to SARS enter the latent class. Latents represent infected who are not yet contagious and are assumed to be asymptomatic, as suggested by results based on epidemiologic, clinical and diagnostic data in Canada [40]. They become infectious after an average time $\varepsilon^{-1}$ (mean latency period). The individual is classified as infectious during an average time equal to $\mu^{-1}$ from the onset of clinical symptoms to his admission to the hospital where he eventually dies or recovers. Patients admitted to the hospital are not allowed to travel. The average periods spent in the hospital from admission to death or recovery are equal to $\mu_D^{-1}$ and $\mu_R^{-1}$, respectively. The average death rate is denoted by $d$.

## Figure 2 - Worldwide map representation of the outbreak likelihood as predicted by the stochastic model

Countries are represented according to the color code, ranging from gray for low outbreak probability to red for high outbreak probability.

## Figure 3 - Map representation of the outbreak likelihood within Canada at the urban area resolution scale

Urban areas are represented according to the color code, ranging from gray for low outbreak probability to red for high outbreak probability. Airports within Canada are also shown.

## Figure 4 - Map representation of the comparison between numerical results and WHO reported cases

Countries are considered at risk if the probability of reporting an outbreak – computed on $n = 10^3$ different realizations of the stochastic noise – is larger than 20%. In red we represent countries for which model forecasts are in agreement with WHO official reports, distinguishing between correct predictions of outbreak (filled red) and correct predictions of no outbreak (striped red). Forecasts that deviate from observed data are represented in green. Results shown refer to the date of 11 July 2003.

## Figure 5 - Number of cases by country: comparison with WHO official reports

Quantitative comparison of forecasted number of cases (conditional of the occurrence of an outbreak) with observed data. Simulated results are represented with a box plot in which lowest and highest values represent the 90% CI and the box is delimited by lower and upper quartile and reports the value of the median. Red symbols represent WHO official reports and are accompanied by the value of the number of cases for sake of clarity. (A) Agreement of model predictions with observed data: symbols are compatible with the model predictions. Broken scale and inset are used for sake of visualization. (B,C) Disagreement of model predictions with observed data: WHO data lie outside the 90% CI obtained from $n = 10^3$ numerical simulations. Results are reported in two different plots characterized by two different scales for a better visualization.

**Figure 6 - Overlap profile**

The value of the overlap is shown as a function of time, from the initial day of the simulations (21 February 2003) to 11 July 2003. Details on relevant events occurring during SARS epidemics are shown for reference.

**Figure 7 - Map representation of epidemic pathways**

Arrows show the paths followed by the virus in the transmission of the infection from Hong Kong to the other countries. The thickness of the arrows represents the probability associated to a given path, where all paths with probability less than 10% have been filtered out for sake of simplicity. Two different colors are used: black for paths that transmit the virus directly from the seed – Hong Kong – to the first level of infected countries; gray for paths that start from the first level of infected countries.

# Tables

## Table 1 - Parameter values

| Parameter | Description | | Baseline value |
|---|---|---|---|
| $T_0$ | Initial offset from 21 February (days) | | 3[*] |
| $\beta$ | Rate of transmission | | 0.57[*] |
| $L(t = 0)$ | Number of initial latent individuals | | 10[*] |
| $s_f(t)$ | Scaling factor for the rate of transmission | 21 February+$T_0$–20 March | 1.00 |
| | | 21 March–9 April | 0.37 |
| | | 10 April–11 July | 0.06 |
| $r_\beta$ | Relative infectiousness of patients at the hospital | | 0.2 |
| $\varepsilon^{-1}$ | Average latency period (days) | | 4.6 |
| $\mu^{-1}(t)$ | Average period from onset of symptoms to admission (days) | 21 February+$T_0$–25 March | 4.84 |
| | | 25 March–1 April | 3.83 |
| | | 2 April–11 July | 3.67 |
| $\mu_R^{-1}$ | Average period from admission to recovery (days) | | 23.5 |
| $\mu_D^{-1}$ | Average period from admission to death (days) | | 35.9 |
| $d$ | Case fatality rate | | 0.2 |

Baseline values for all epidemiological parameters and initial conditions. Parameters marked with an asterisk ([*]) are estimated by our model through the fitting procedure described in the main text. The three successive decreasing values for the $\mu^{-1}$ model are the prompter, identification and subsequent isolation of infectious individuals [11]. A step function is also assumed for the scaling factor $s_f(t)$ of the transmission parameter $\beta$, with values taken from the estimates of the effective reproductive number $R_t$ with respect to $R_0$ during the early stage of SARS epidemic in Hong Kong [9]. This corresponds to the effective reduction of the reproductive number due to the application of control measures [9].

**Table 2 - Forecasted number of cases for the countries with an incorrect prediction of outbreak**

| Country | Median | 90% CI |
|---|---|---|
| Japan | 83 | 23−228 |
| United Arab Emirates | 6 | 1−36 |
| Bangladesh | 6 | 1−42 |
| Saudi Arabia | 5 | 1−35 |
| Netherlands | 5 | 1−26 |
| Cambodia | 5 | 1−40 |
| Bahrain | 4 | 1−30 |
| Austria | 4 | 1−26 |
| Denmark | 3 | 1−15 |
| Brunei | 3 | 1−16 |

List of countries that were not infected according to WHO official reports but are predicted as at risk by numerical simulations. Median and 90% CI are reported; results correspond to 11 July 2003.

# Figures

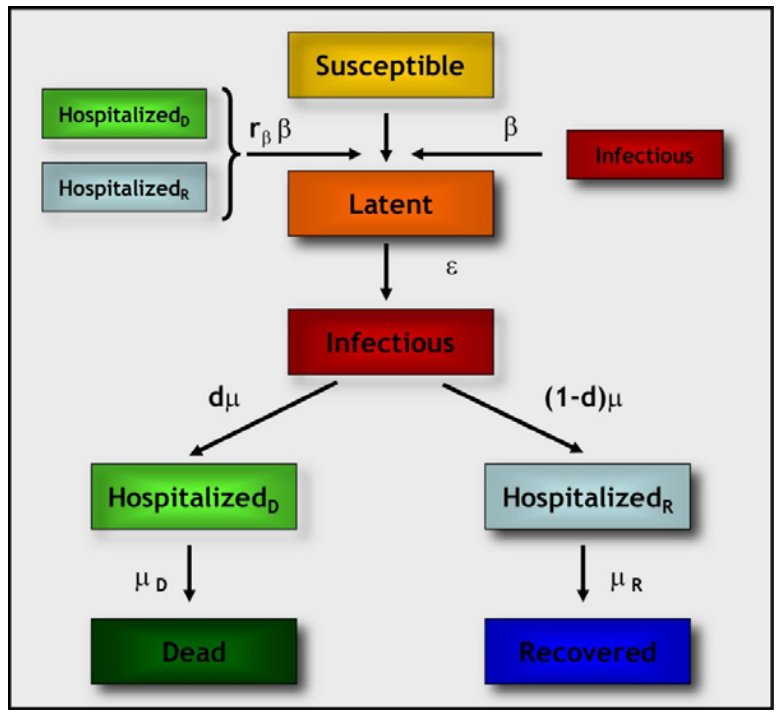

Figure 1.

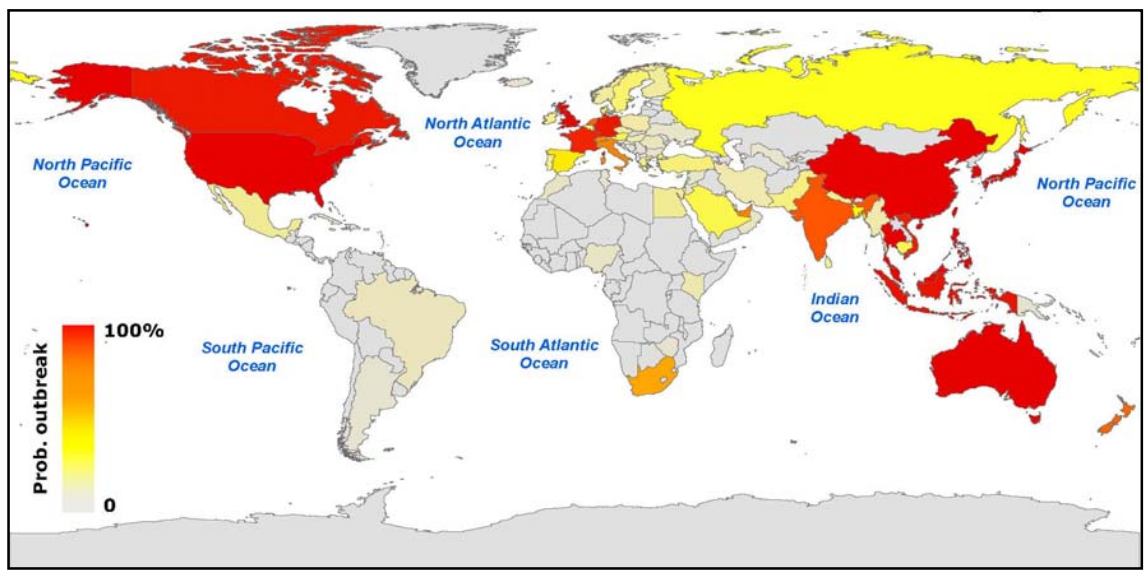

Figure 2.

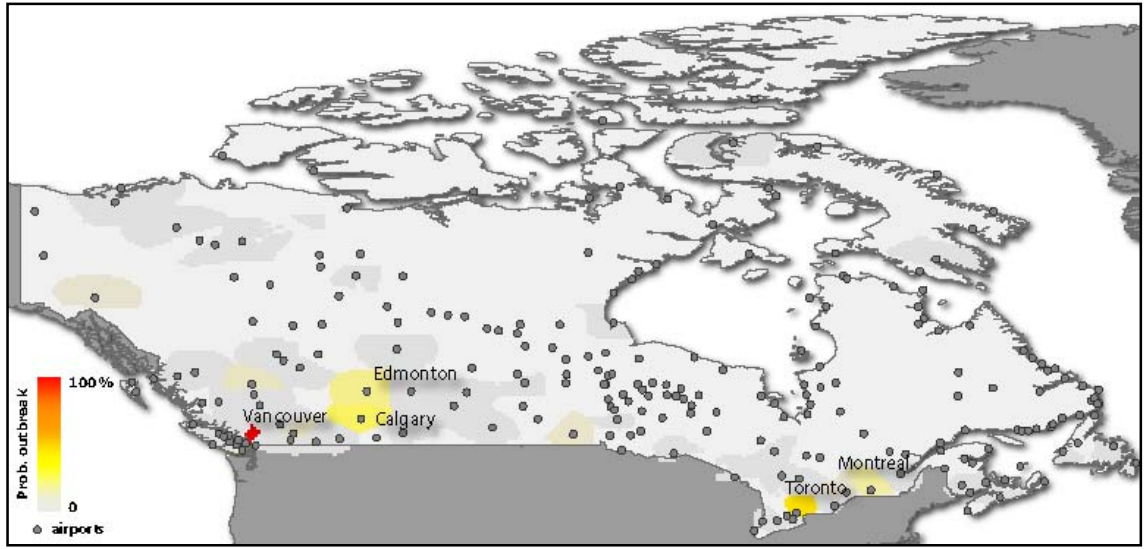

Figure 3.

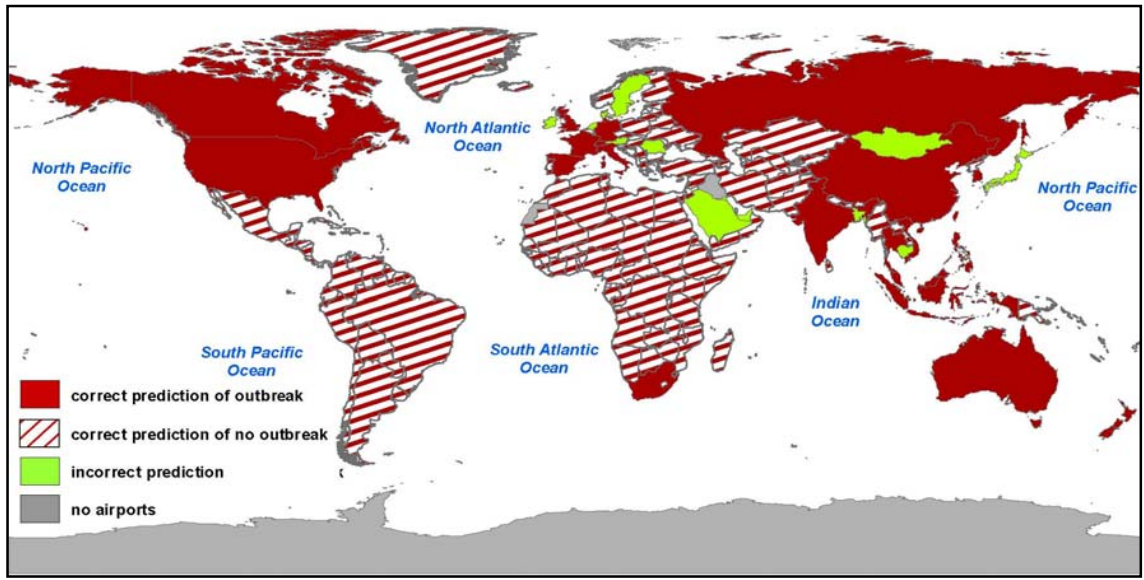

Figure 4.

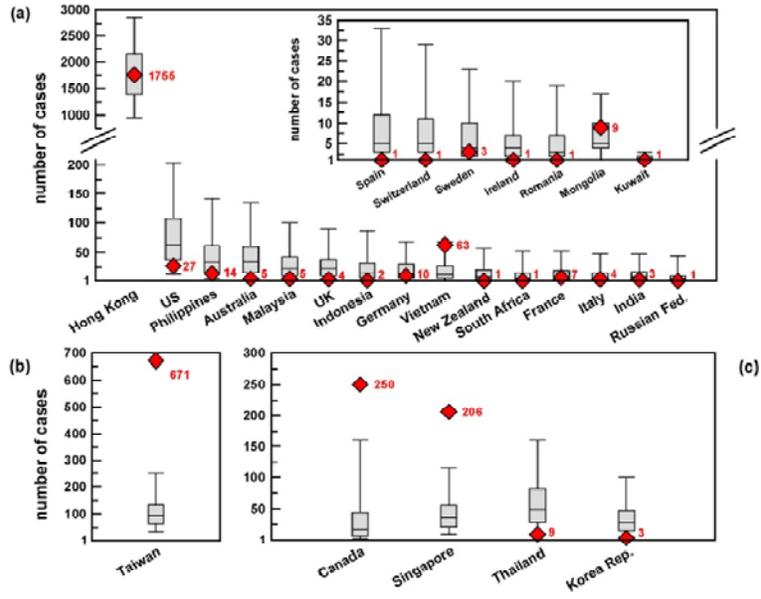

Figure 5.

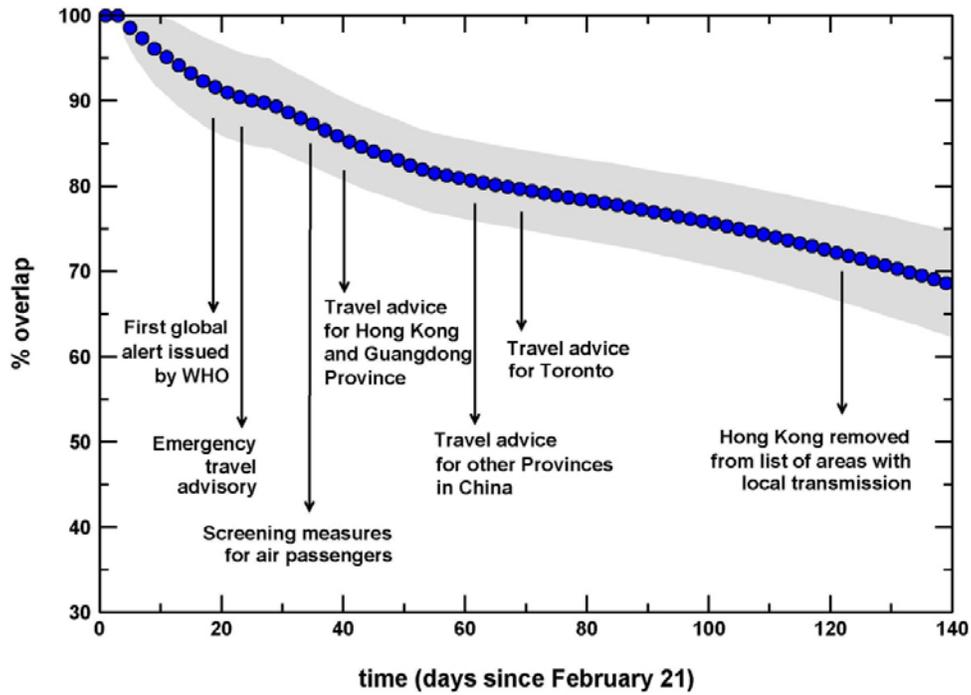

Figure 6

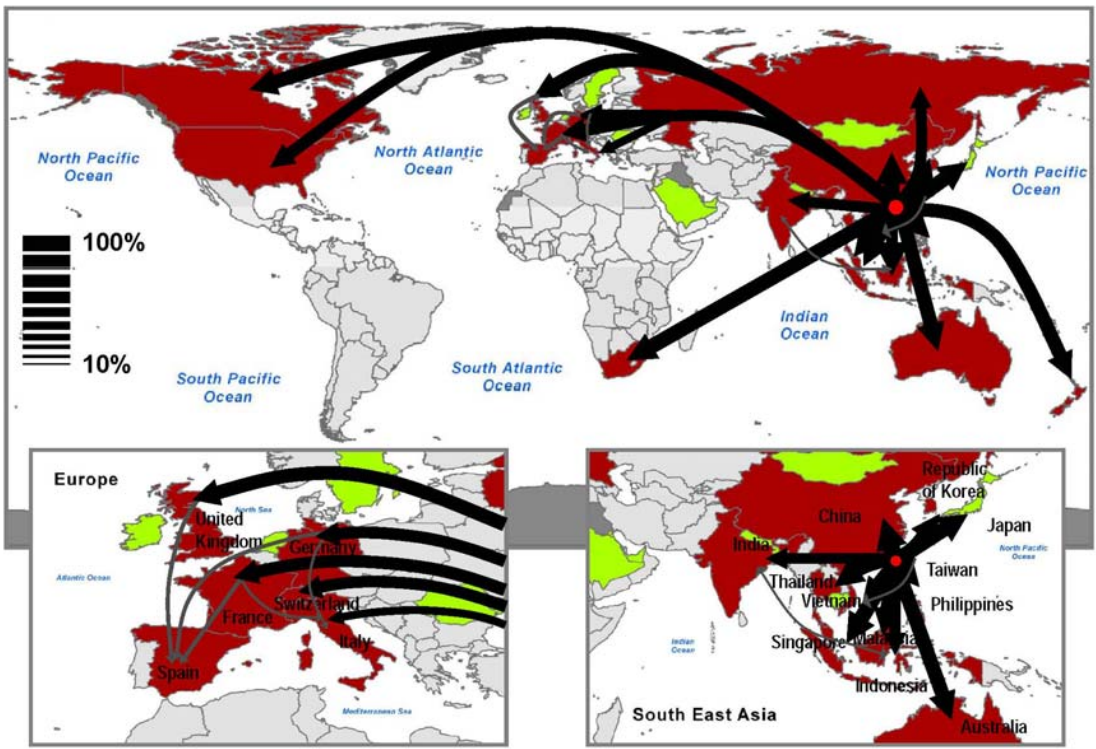

Figure 7.